\documentstyle[prb,twocolumn,aps]{revtex}
\input{psfig}

\preprint{}
\begin{document}
\draft
\title{ac Josephson effect in superconducting $d$-wave junctions}
\author{Magnus Hurd\cite{hurd_address}, Tomas L\"ofwander,
G\"oran Johansson, and G\"oran Wendin}
\address{Department of Microelectronics and Nanoscience,
School of Physics and Engineering Physics,\\
Chalmers University of Technology and G\"oteborg University,\\
S-412 96 G\"oteborg, Sweden}
\date{\today}
\maketitle
\begin{abstract}
  
  We study theoretically the ac Josephson effect in superconducting
  planar $d$-wave junctions. The insulating barrier assumed to be
  present between the two superconductors may have arbitrary strength.
  Many properties of this system depend on the orientation of the
  $d$-wave superconductor: we calculate the ac components of the
  Josephson current. In some arrangements there is substantial
  negative differential conductance due to the presence of mid-gap
  states.  We study how robust these features are to finite
  temperature and also comment on how the calculated current-voltage
  curves compare with experiments. For some other configurations (for
  small barrier strength) we find zero-bias conductance peaks due to
  multiple Andreev reflections through midgap states. Moreover, the
  odd ac components are strongly suppressed and even absent in some
  arrangements.  This absence will lead to a doubling of the Josephson
  frequency.  All these features are due to the $d$-wave order
  parameter changing sign when rotated $90^{\circ}$. Recently, there
  have been several theoretical reports on parallel current in the
  $d$-wave case for both the stationary Josephson junction and for the
  normal metal-superconductor junction.  Also in our case there may
  appear current density parallel to the junction, and we present a
  few examples when this takes place. Finally, we give a fairly
  complete account of the method used and also discuss how numerical
  calculations should be performed in order to produce current-voltage
  curves.

\end{abstract}

 
\narrowtext

\section{Introduction}\label{Sec_1}
There are now strong reasons to believe that many of the high-$T_c$
superconductors (HTS) exhibit a superconducting order parameter which
has mainly (or completely) $d_{x^2-y^2}$-wave ($d$-wave)
symmetry.\cite{Woll,Harl} The exact structure of the order parameter
depends on the material. For instance, due to the orthorhombic
character of $\mbox{Y}\mbox{Ba}_2 \mbox{Cu}_3\mbox{O}_{7-\delta}$
(YBCO) an $s$-wave component of the order parameter is
induced.\cite{Kouz} This is not the case for tetragonal systems like
$\mbox{Tl}_2\mbox{Ba}_2 \mbox{Cu}\mbox{O}_{6+\delta}$, where a pure
$d_{x^2-y^2}$-wave symmetry has been established.\cite{Tsue3} At any
rate, the presence of a substantial component of $d$-wave symmetry in
the high-$T_c$ superconductors cannot be ignored.
  
This paper sets out to explore theoretically the implications of
$d$-wave symmetry for the ac Josephson effect in high-$T_c$
superconductors.

\begin{figure}
\centerline{\psfig{figure=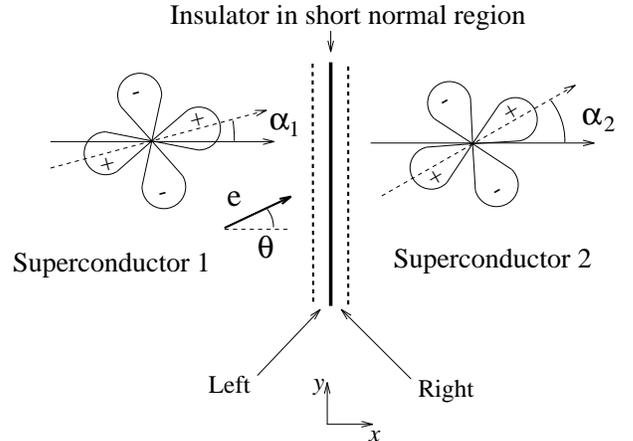,width=8cm}}
\caption{Layout of the junction. The orientations of the two
  $d$-wave superconductors are described by the angles $\alpha_1$ and
  $\alpha_2$. An electron-like quasiparticle is incident on the
  junction at an angle $\theta$. Short normal regions are introduced
  between the barrier and the superconductors. The strength of the
  barrier can be tuned from the ballistic to the insulating case.}
\label{fig1}
\end{figure}

To have the ac Josephson effect, two superconductors must be present
on both sides of a normal (or insulating) region as shown in
Fig.~\ref{fig1}. For an anisotropic superconductor the order parameter
$\Delta(\theta)$ depends on the angle $\theta$ of incidence of the
quasiparticle approaching the junction region.  In the case of pure
$d$-wave symmetry we have $\Delta(\theta)=\Delta_0
\cos[2(\theta-\alpha)]$, where $\alpha$ is the orientation of the
$d$-wave order parameter with respect to the insulator. We will in
this paper use the notation $d_\alpha$ for $d$-wave superconductor
with orientation $\alpha$. For the isotropic $s$-wave case,
$\Delta(\theta)=\Delta_s$. Another competing candidate for the order
parameter of HTS has been the anisotropic $s$-wave order parameter:
$\Delta(\theta)=\Delta_0 \cos^4[2(\theta-\alpha)]+\Delta_1$; today,
however, it does not seem to be on the short list of possible HTS
order parameters any longer.\cite{Harl,Kouz,Tsue3}

The strength of the insulating barrier shown in Fig.~\ref{fig1} may
vary in our treatment.  This makes it possible to treat all
superconductor-superconductor junctions from the ballistic (fully
transparent) superconductor-normal metal-superconductor (SNS) junction
to the superconductor-insulator-superconductor (SIS) tunnel junction.
To describe superconductor-superconductor junctions in general we use
the notation $\mbox{S}_1|\mbox{S}_2$ where $\mbox{S}_{1(2)}$ is the
superconductor on the left (right) side of the interface $|$. The
interface is a barrier of arbitrary strength.

The ac Josephson effect, from a general point of view, means that for
an $\mbox{S}|\mbox{S}$ junction a time independent bias voltage $V$
results in a time dependent current. The non-stationary current
perpendicular to the junction (the $x$-direction; see Fig.~\ref{fig1})
can be decomposed into frequency components
\begin{equation}
I_x(V,t)=\sum_m I_{x,m}(V) e^{i m \omega_J t} ,
\label{ac_curr_x}
\end{equation}
where $m$ is an integer. The Josephson frequency $\omega_J= 2 e
V/\hbar$ corresponds to the energy gained by a Cooper pair when
passing through the junction region. Since the relation
$I_{x,-m}(V)=I^*_{x,m}(V)$ holds, the current is real.

An order parameter with $d_{x^2-y^2}$-wave symmetry changes sign when
rotated $90^{\circ}$. This feature has been shown to have several
implications on transport properties of $d_{x^2-y^2}$-wave
superconductors, some of which have been confirmed in experiments:
cancellation effects as to the dc Josephson
current\cite{Woll,Harl,Kouz,Tsue3} and the presence of zero-bias
conductance peaks (ZBCP) in various structures of HTS
junctions.\cite{GXL,Cov,SinNg,Alff1,Alff2,Alff3} Since in this paper
the focus is on the ac Josephson effect we will (among other things)
discuss issues related to ZBCP.

The ZBCP arises due to the presence of zero-energy bound states,
so-called midgap states (MGS), at surfaces/interfaces of
$d_{x^2-y^2}$-wave superconductors.\cite{Hu} The reason MGS appear is
an interplay between normal scattering (at the surface/interface) and
the fact that the $d$-wave order parameter changes sign at some
regions in momentum space. Since in the $d$-wave case the order
parameter depends on momentum (or in other words angle $\theta$), a
quasiparticle will in general sense a different order parameter before
and after scattering (normal scattering changes the momentum). When
there is a sign difference of the order parameter before and after
scattering a zero-energy bound state appears (MGS).\cite{Hu} These
localized zero-energy states open up additional channels for current
flow, leading to peaks in the conductance.  Calculations of the
current-voltage ($I-V$) curves for the normal metal-superconductor
(NS) junction\cite{TanKas} confirmed the picture brought forward in
Ref.~\onlinecite{Hu}. MGS are infinitely sharp when the transmission
of the strength of the barrier is infinetely high.  When the strength
of the barrier decreases the MGS broaden.

In addition, there are some other features related to the ac Josephson
effect involving $d$-wave superconductors that have not yet (to our
knowledge) been experimentally established, such as negative
differential conductance\cite{Hurd,BarSvi1,BarSvi2,Wendin} and
cancellation of the odd ac components, doubling the Josephson
frequency,\cite{LJHW} although there are experiments discussing
related features.\cite{SinNg,Char} Negative differential conductance
is a result of resonant conduction through MGS.\cite{Hurd} Moreover,
we will in this paper show an example of ZBCP for the case when there
are MGS on both sides of the barrier (possible only when there are
$d$-wave superconductors on both sides of the barrier). We note that
for the cases we discuss ZBCP appear for relatively small values of
barrier strength. The geometry used in these calculations correspond
to experiments just recently performed.\cite{Alff2,Alff3}

\begin{figure}
\centerline{\psfig{figure=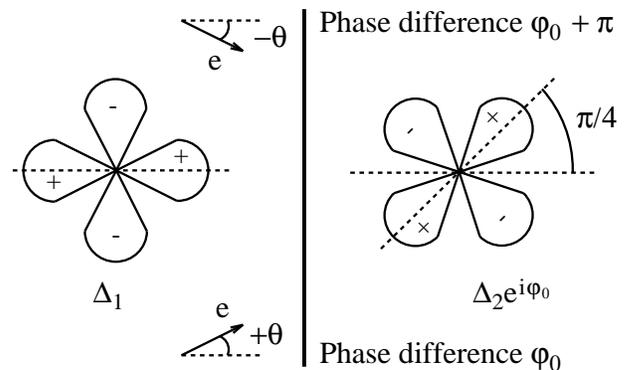,width=8cm}}
\caption{Symmetry argument for the cancellation of the odd ac
  components for the specific orientation $d_{0}|d_{\pi/4}$.
  Considering injection of electron-like quasiparticles at angles
  $\pm\theta$, the gap in the left superconductor will be the same,
  while there will be a sign difference in the right superconductor.
  In the above figure, the negative sign appears at $-\theta$ and can
  be thought of as an extra phase $\pi$ for this injection angle.  The
  current contributions from $\pm\theta$ are equal in magnitude but
  have different signs for the odd components, because of the extra
  phase $\pi$, and will therefore cancel.}
\label{fig2}
\end{figure}

Cancellation of the odd ac components is explained as follows (see
Fig.~\ref{fig2}). The keypoint is that the total current is a sum of
contributions, associated with different angles $\theta$.  In the case
of two $d$-wave superconductors with orientation angles $\alpha_1=0$
and $\alpha_2=\pi/4$ respectively, quasiparticles incident at angle
$-\theta$ experience a different sign of the gap $\Delta_2$ compared
to quasiparticles incident at $\theta$. This sign difference (which
corresponds to a phase difference of $\pi$) enters the expression of
ac components as $e^{i m \pi}=(-1)^m$, where $m$ is the index of the
ac component in Eq.~(\ref{ac_curr_x}). This means that for even (odd)
$m$, positive and negative angles add up (cancel out).\cite{LJHW} We
note that the doubling of the Josephson frequency is completely in
line with the change of the current-phase relationship in stationary
Josephson junctions from being $2\pi$-periodic to being $\pi$-periodic
in some arrangements.\cite{Yip1,Yip2,TanKas2,Zago}

Moreover, there appears in voltage-biased junctions of $d$-wave
superconductors a possibility to have current flow parallel to the
junction. This is not necessarily related to the ac Josephson effect
but could appear in NS junctions as well.\cite{LJHW}

Recently, there has been a discussion concerning the appearance of
superconducting states breaking time-reversal symmetry, especially at
inhomogeneities (like surfaces/interfaces) in $d$-wave
superconductors. These states have been theoretically predicted to
have an order parameter of the form $d+is$, where the $is$-part is
induced by the inhomogeneity and only present close to the
inhomogeneity. The $is$-wave part of the order parameter has vital
physical importance: the $d+is$ combination explicitly breaks
time-reversal symmetry, and peculiar surface/interface currents appear
with the existence of the $is$ part. One example of this phenomenon is
twin boundaries of YBCO, whose order parameter is of the form $d \pm
s$ (a real combination). The twin boundary divides the superconductor
into two parts with different signs in front of the $s$-part. To match
the two sides the most favorable way (from a Ginzburg-Landau
theoretical point of view), close to the boundary a complex order
parameter appears in order to avoid a discontinous change in the phase
between the $s$- and $d$-part of the order parameter.\cite{Sig} This
result has also been found within a Green's function
formalism\cite{BBS} and by solving the Bogoliubov-de Gennes
equation.\cite{Feder} Another example is the surface of a $d$-wave
superconductor, where the surface lowers the symmetry in such a way
that a combination of the subdominant $s$ and the dominant $d$ order
parameters are now allowed at temperatures below a certain temperature
defined by various properties, giving rise to a $d+is$ surface state
with associated time-reversal symmetry breaking currents parallel to
the surface.\cite{Cov,MatShi,BPRS,FRS} These surface states are doubly
degenerate, corresponding to two opposite current directions. A third
example (and most closely related to the case studied in this paper)
is the junction between an $s$-wave superconductor and a
$d_{x^2-y^2}$-wave superconductor rotated $45^{\circ}$,\cite{Sigrist}
where a combination of $s$- and $d$-waves coexists close to the
junction area.  Again the phase difference of the energetically stable
phase is such that a $d\pm is$ order parameter appears, also this time
doubly degenerate corresponding to phase differences of $\pm \pi/2$
between the $s$- and $d$-wave side.  The same effect has theoretically
been found in junctions between two $d_{x^2-y^2}$ superconductors
where one of the superconductors is rotated by $45^o$.\cite{FY}

\begin{figure}
\centerline{\psfig{figure=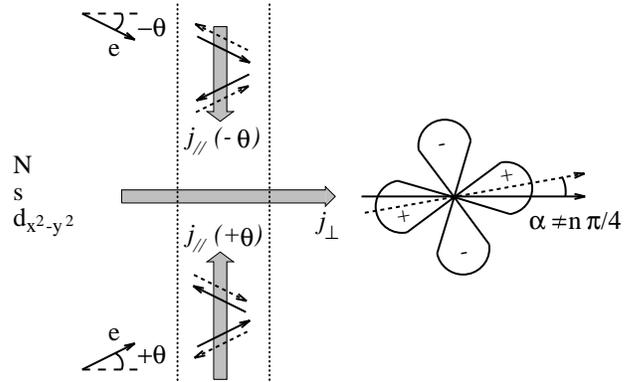,width=8cm}}
\caption{This picture shows how the parallel ($y$-direction)
  current appears as a consequence of the asymmetry of the $d$-wave
  order parameter.  On the left side we have a normal, an $s$-wave or
  a $d$-wave structure; on the right side we have a $d$-wave
  superconductor with $\alpha\neq n \pi/4$, where $n$ is an integer.
  Quasiparticles incident on the right side (from the left) in general
  experience a different gap depending on angle of incidence $\theta$.
  Since current contributions (among other things) depend on the gap
  this means that the parallel current contribution corresponding to
  $\theta$ does not necessarily cancel the one corresponding to
  $-\theta$. The solid and dashed lines illustrate electrons and holes
  propagating in both directions.}
\label{fig3}
\end{figure}

All the examples discussed above concern the stationary state.  This
is definitely different from our case with a voltage present between
the two sides of the junction.\cite{Hurd,LJHW} In fact, when the
junction is voltage-biased it does not care about the ground state
phase difference between the two sides. Rather, all possible phases
are taking part in an averaging procedure. Therefore, it is not
surprising that the case discussed in Ref.~\onlinecite{Sigrist} (the
third example in the previous paragraph) does not produce parallel
current density for the zero-frequency ac component in the
voltage-biased case, since the two degenerate states (corresponding to
two opposite current directions) are now averaged to produce a zero
net dc current density. In Fig.~\ref{fig3} we show the mechanism in
the voltage biased case for the net current density parallel to the
interface between the two sides.  The bottomline is that a
$d_{x^2-y^2}$-wave order parameter with orientation angle $\alpha \neq
0$ treats quasiparticles associated with positive injection angles
differently than quasiparticles associated with negative injection
angles, leading to an imbalance between current contributions
associated with positive and negative injection angles.

Therefore, in this paper we consider the ac Josephson effect and focus
on three main issues: first, current-voltage curves of $d$-wave
$\mbox{S}_1|\mbox{S}_2$ junctions; second, the disappearance of the
odd ac components; third, the current density parallel to the junction
direction. Moreover, we give a fairly complete account of the methods
used.\cite{Hurd,LJHW} Especially, we elaborate on how various ways of
writing down current formulas relate to each other.
 
In the case of $s$-wave superconductivity Eq.~(\ref{ac_curr_x}) has
been studied for a long time. In the tunneling limit (low transmission
of the insulating barrier) detailed calculations\cite{Fal,Jose,AmbBar}
showed excellent agreement with experimental results for the
conventional $s$-wave superconductors. These calculations used the
tunneling hamiltonian method. Due to the huge success of the tunneling
calculations less attention was paid to the non-tunneling case.

The work by Blonder, Tinkham, and Klapwijk proved that with fairly
simple methods it was possible to produce current-voltage curves of NS
junctions for any barrier strength.\cite{BTK} Later, another line of
research focused on introducing the concept of mesoscopics into
superconductivity.\cite{BeenHout,Been,GunZai,BSW,AveBar} Both the dc
Josephson effect\cite{BeenHout,Been} and the ac Josephson
effect\cite{GunZai,BSW,AveBar} were studied, again without the
tunneling-limit restriction.

The main tool in many of these papers dealing with the case of
arbitrary transmission has been the Bogoliubov-de Gennes (BdG)
equation.\cite{BTK,BeenHout,Been,BSW,AveBar} The time-dependent BdG
equation together with a formula for the current gives a complete
description of the current in both the stationary and non-stationary
case. In this paper we use a generalization\cite{Hurd,LJHW} of the
methods worked out for the $s$-wave case.\cite{BSW,AveBar,SBW,BSBW}
 
When an electron with subgap energy from a normal region approaches
the superconductor, there seems to be no way of transmitting electric
charge into the superconductor: there are no propagating states of the
superconductor in the subgap region. This becomes evident when solving
the Bogoliubov-de Gennes (BdG) equation resulting in decaying wave
functions when solving for subgap energies.\cite{BTK,deG}

\begin{figure}
\centerline{\psfig{figure=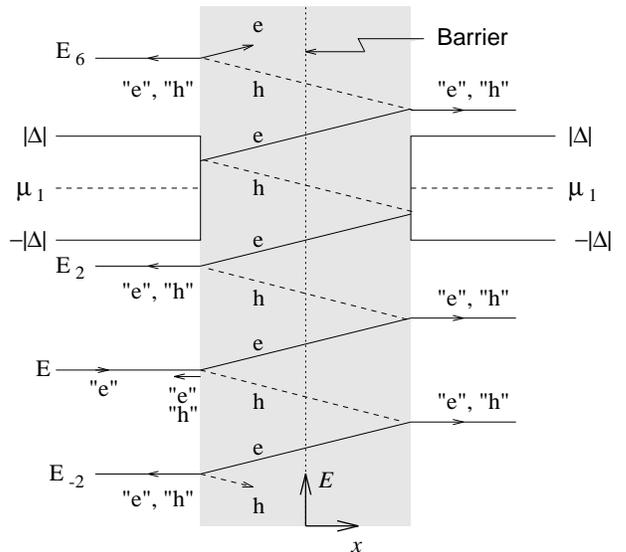,width=8cm}}
\caption{The scattering state originating from an electron-like
  quasiparticle incident at energy $E$. There are electrons and holes
  propagating in both directions (no arrows included in the figure) in
  the normal region at energies $E_n=E+neV$, where $n$ is an integer
  and $V$ is the voltage over the junction. The shaded area is the
  junction region where the voltage drop occurs.}
\label{fig4}
\end{figure}

However, there is one important physical process making transmission
of electric charge possible also for subgap incidence: Andreev
reflection (AR).\cite{BTK,And} In this process an electron (from a
normal region) with subgap energy incident on a superconductor is
reflected as a hole moving in the opposite direction to the original
electron. This event does obviously not in itself conserve particles.
Therefore, a Cooper pair of electrons is injected into the
superconductor. When two superconductors embrace the normal region,
forming an SNS junction, the hole will finally hit the other
superconductor and again Andreev reflects, this time as an electron.
These AR's will continue back and forth in the normal region.  Since
the voltage drop is over the normal region, the electrons and holes
gain the energy $eV$ every time they pass the normal region (see
Fig.~\ref{fig4}). These repeated reflections are usually named
multiple Andreev reflections (MAR). The reflections are perfect inside
the gap but decrease in strength when the magnitude of the excitation
energy is increased away from the gap region.

For each round trip two AR's transform the original electron into an
electron at an energy $2eV$ higher than from the start. This means
that the wave functions contain parts (sidebands) separated by
multiples of $2eV$ from each other, leading to an ac current with a
frequency of $2eV/\hbar$ as seen in Eq.~(\ref{ac_curr_x}).

We think Fig.~\ref{fig4} is a key figure in order to illustrate the
underlying physical mechanism of the phenomenon dicussed in this paper
(the ac Josephson effect): it makes clear the meaning of
Eq.~(\ref{ac_curr_x}). There are two scattering events of our problem:
first, normal scattering at the barrier described by a transmission
amplitude; second, AR at the NS interfaces described by an AR
amplitude. Therefore, Eq.~(\ref{ac_curr_x}) can be thought of as an
expansion in transmission amplitude (through the barrier) and AR
amplitude. Higher orders of the wave function corresponding to
sidebands visualized in Fig.~\ref{fig4} contain higher orders of
transmission through the barrier and higher orders of AR.

The present paper is organized in the following way. In
Section~\ref{Sec_2} we outline the model, which is used in
Section~\ref{Sec_3} to calculate the various ac components of the
current. We present the results in Section~\ref{Sec_4}, and conclude
with a summary in Section~\ref{Sec_5}. We feel that an important part
of our paper is given in the appendices: in Appendix~\ref{App_A} the
time-dependent BdG equation is solved, and we also give some numerical
advice on how to calculate the current; in Appendix~\ref{App_B} we
compare how various current formulas previously used in the literature
relate to each other; in Appendix~\ref{App_C} we point out a couple of
symmetry relations that are relevant to our problem; finally, in
Appendix~\ref{App_D} we write down a pair of analytical expressions
describing the ballistic case.

\section{The model}\label{Sec_2}
To mathematically describe the voltage-biased $\mbox{S}_1|\mbox{S}_2$
system shown in Fig.~\ref{fig1} we have made a number of assumptions,
some of them pertaining to the superconducting parts of the system and
some of them pertaining to the normal part of the system. In the
following we detail these assumptions.

Starting with the normal part (the junction) in between the
superconductors, we assume a planar structure with translational
invariance in the $y$-direction. The planar geometry is a necessary
condition for MGS to exist, contrary to the case of point
contacts\cite{HDB2} where no MGS are around.  Moreover, we neglect in
our treatment any influence from interface roughness. In the normal
region we only take into account scattering due to the barrier, which
is modeled as a one-particle potential $V(x)=H\delta(x)$. The
$\delta$-function form of $V(x)$ results in a reflection amplitude
$r(\theta)=Z/(i\cos\theta-Z)$ and a transmission amplitude
$t(\theta)=i\cos\theta/(i\cos\theta-Z)$, where $Z$ is defined as
$Z=2mH/\hbar^2$.\cite{Brud} The parameter $Z$ describes the strength
of the barrier: $Z=0$ corresponds to no scattering at all (ballistic
case); $Z=\infty$ corresponds to the case when the two sides of the
junction are uncoupled (no $x$-current in this case); a high value of
$Z$ corresponds to the tunneling case.

Next we consider the superconducting parts. In our calculation
self-consistency is not fulfilled. In the literature there is work
that has included self-consistency in the tunneling limit showing that
non-zero energy states appear along with MGS.\cite{BarSvi1,BarSvi2} In
our work we neglect these effects; on the other hand, we allow the
barrier strength $Z$ of the barrier to have any value.

Therefore, we assume that the superconducting order parameter of the
structure in Fig.~\ref{fig1} is
\begin{equation}
\Delta=\left\{
\begin{array}{ll}
\Delta_1(\theta), & x<0\nonumber\\
\Delta_2(\theta)e^{i \phi_0}, & x>0.
\end{array}
\right.
\label{order_parameter}
\end{equation}
Since the overall phase is unimportant we can here choose $\Delta_1$
real. There may be a phase-difference $\phi_0$ over the junction which
we defer to the right-hand superconductor. The angle $\theta$
corresponds to the angle of incidence of the quasiparticle approaching
the superconductor. In the following we will use various forms of
order parameters for $\Delta_1(\theta)$ and $\Delta_2(\theta)$ ranging
from the $s$-wave case to the $d$-wave case.  The $d$-wave
superconductors are allowed to be rotated an angle $\alpha$ measured
relative to the interface normal as described in Fig.~\ref{fig1}.

In this paper we do not include the magnetic field. Including the
magnetic field will screen any current on the length scale of the
London penetration depth, forcing the current to flow next to
surfaces/interfaces of superconductors. Neglecting screening effects
is traditionally considered valid if the system under study has
dimensions smaller than the London penetration depth.  However, a
finite system immediately raises questions about where the parallel
current finally ends up. In this paper we do not address this
question; instead we just note that a full treatment including
screening effects will be required to understand how the parallel
current is guided close to the interface of the $d$-wave
superconductor.

Using the models discussed above for both the normal and the
superconducting regions we solve the time-dependent Bogoliubov-de
Gennes (BdG) equation for anisotropic superconductors
\begin{eqnarray}
&&\int d^3x'\left(
\begin{array}{cc}
h_0({\bf x},{\bf x}') & \Delta({\bf x},{\bf x}',t)\\
\Delta^*({\bf x},{\bf x}',t) & -h_0({\bf x},{\bf x}')
\end{array}
\right)\left(
\begin{array}{c}
u({\bf x}',t)\\
v({\bf x}',t)
\end{array}
\right)=
\nonumber\\
&&i\hbar\partial_t\left(
\begin{array}{c}
u({\bf x},t)\\
v({\bf x},t)
\end{array}
\right),\ \
h_0({\bf x},{\bf x}')=\delta({\bf x}-{\bf
x}')\left(\frac{p_{x'}^2}{2m}-\mu\right),
\label{BdG}
\end{eqnarray}
where $\mu$ is the chemical potential, and $\Delta({\bf x},{\bf
  x}',t)$ is the order parameter. Introducing a voltage the chemical
potentials of the two reservoirs will be shifted relative to each
other, $eV=\mu_2-\mu_1$. The natural energy reference level in a
superconductor is the chemical potential. In order to establish a
global energy reference level, a gauge transformation is performed as
described in detail in Ref.~\onlinecite{HDB1}. As a result of this
transformation a phase difference appears over the junction obeying
the Josephson relation $\partial_t\phi=2eV/\hbar$. In other words, for
our case with constant voltage we describe this time-dependent phase
difference multiplying $\Delta_2(\theta)$ by $\exp(i2eVt/\hbar)$.

In Appendix~\ref{App_A} we solve Eq.~(\ref{BdG}) for the wave function
$\Psi_{\nu}=(u_{\nu},v_{\nu})$ corresponding to an electron-like
quasiparticle incident on the junction at an energy $E_{\nu}$.  Using
this wave function we will in the next section calculate the current.

\section{Calculating the current}\label{Sec_3}
To calculate the current density ${\bf j}$ we use the general formula
\begin{equation}
{\bf j}=\frac{e\hbar}{m}\sum_{\nu} \left[2f(E_{\nu})-1\right]
\mbox{Im}\left\{u_{\nu}^*\nabla u_{\nu}+v_{\nu}^*\nabla v_{\nu}
\right\},
\label{currentdensity}
\end{equation}
where we introduced the Fermi distribution function
$f(E)=1/[1+\exp(E/k_BT)]$.  The sum is over all scattering states
originating from electron-like quasiparticles approaching the junction
from the superconducting reservoirs at both negative and positive
energies $E_{\nu}$. There are different ways of writing the formula
for the current density, but as shown in Appendix~\ref{App_B} they are
all equivalent.  Note that $\bf j$ in our case may have a component in
the $y$-direction (parallel to the junction). The wave function
coefficients determined in Appendix~\ref{App_A} are introduced into
the current formula of Eq.~(\ref{currentdensity}) in a way outlined
below.

\subsection{The $x$-current}

We start by studying the current density $j_x$ perpendicular to the
junction. Since we consider a structure smaller than the London
penetration depth (allowing us to neglect self-field effects) the
current density will distribute uniformly over the junction.
Therefore we may calculate the current density at any $y$-coordinate.
Moreover, a conservation law for the current density assures that we
may calculate $j_x$ at any $x$-coordinate; it proves convenient to
calculate the current density in the normal region (in our treatment
to the left of the barrier in Fig.~\ref{fig1}) between the
superconducting reservoirs. The total perpendicular current $I_x$ per
$ab$-plane is then given by $I_x=L_y j_x$, where $L_y$ is the length
of the junction in the $y$-direction.

In Eq.~(\ref{currentdensity}) the quantum number $\nu$ is the wave
vector ${\bf k}$ of an electron-like quasiparticle approaching the
junction.  The direction could be thought of as another quantum
number; we introduce $\tau=+$ ($-$) to indicate that the quasiparticle
is incident from the left (right). Therefore, using the wave function
coefficients of Appendix \ref{App_A}, the current per $ab$-plane is
\begin{equation}
I_x=L_y \frac{e\hbar}{m} \frac{k_F}{L_x L_y} \sum_{{\bf k},\tau}
\tau 
\cos(\theta_{\bf k}) T^{\tau}({\bf k}) [2f(E_{\bf k})-1],
\label{curr_1}
\end{equation}
where the combination of $k_F$ (the Fermi wavevector) and
$\cos(\theta_{\bf k})$ ($\theta_{\bf k}$ is the angle of incidence for
an electron-like quasiparticle incident with momentum $\bf k$) is
produced by the $x$-derivative in Eq.~(\ref{currentdensity}). The
factor $1/(L_x L_y)$ in the equation above normalizes the wavefunction
of the incoming electronlike quasiparticle. In Eq.~(\ref{curr_1}) we
have introduced
\begin{eqnarray}
T^{\tau}({\bf k})&=&\mbox{Re}\sum_{n,m}
\left[(a_{L,2n+2m}^{\tau})^* a_{L,2n}^{\tau}
-(d_{L,2n+2m}^{\tau})^* d_{L,2n}^{\tau}\right.\nonumber\\
&&\left.+(b_{L,2n+2m}^{\tau})^* b_{L,2n}^{\tau}
-(c_{L,2n+2m}^{\tau})^* c_{L,2n}^{\tau}\right]\times \nonumber\\
&&e^{i\tau m2eVt/\hbar}e^{i\tau m\phi_0}.
\label{transmission}
\end{eqnarray}
The set of states (electron-like quasiparticles) to sum over in
Eq.~(\ref{curr_1}) is illustrated in Fig.~\ref{fig11}(c). In
Eq.~(\ref{transmission}) $n$ and $m$ are integer numbers.

Next we change the sum over $\bf k$ in Eq.~(\ref{curr_1}) into an
integration over $d^2 k$.  This produces a factor of $L_x
L_y/(2\pi)^2$. Finally, changing the integration over $d^2 k$ into an
integration over the superconducting energy $E=[(\hbar^2
k^2/2m-\mu)^2+\Delta(\theta_{\bf k})^2]^{1/2}$ and the angle of
incidence $\theta=\theta_{\bf k}$, we find the following expression
for the current
\begin{eqnarray}
\frac{I_x}{\sigma_0}=&&\frac{eV}{\Delta_0}+
\frac{1}{4D}\int_{-\pi/2}^{\pi/2}d\theta\cos\theta
\int_{-\infty}^{\infty}\frac{dE}{\Delta_0}[2f(E)-1]
\nonumber\\
&&\times\sum_{\tau} \tau N_\tau(E,\theta)T^{\tau}(E,\theta),
\nonumber\\
\sigma_0=&&L_y\frac{2^{5/2}em^{1/2}E_{F}^{1/2}\Delta_0D}{h^2}\; ,
\label{curr}
\end{eqnarray}
where $D=\int d\theta |t(\theta)|^2 \cos\theta/2$ and $E_F$ is the
Fermi energy.  The first term proportional to $eV$ in Eq.~(\ref{curr})
is due to refering energies of right-movers (left-movers) to the
chemical potential of the left (right) superconductors. For details,
see Refs.~\onlinecite{HDB2} and \onlinecite{HDB1}. We note that the
second term of Eq.~(\ref{curr}) is zero when all order parameters are
put to zero. It is therefore reasonable to call this part the
superconducting contribution to the current and the first part the
normal contribution to the current.

It is now convenient to introduce $\Omega_m=m2eVt/\hbar+m\phi_0$.  We
then rewrite Eq.~(\ref{curr})
\begin{equation}
\frac{I_x}{\sigma_0}=\frac{eV}{\Delta_0}+
\sum_m[A_{x,m}\cos\Omega_m+B_{x,m}\sin\Omega_m],
\label{Isincos}
\end{equation}
where $A_{x,m}$ and $B_{x,m}$ are given by
\begin{eqnarray}
A_{x,m}(V)=&&\frac{1}{4D}\int_{-\pi/2}^{\pi/2}d\theta \cos\theta\
\int_{-\infty}^{\infty}\frac{dE}{\Delta_0}
\tanh\left(\frac{-E}{2k_B T}\right)\nonumber\\
&&\times\sum_\tau \tau N_\tau (E){\mbox{Re}}\left\{
T_x^{\tau}(E,\theta,m)\right\},\nonumber\\
B_{x,m}(V)=&&-\frac{1}{4D}\int_{-\pi/2}^{\pi/2}d\theta \cos\theta
\int_{-\infty}^{\infty}\frac{dE}{\Delta_0}
\tanh\left(\frac{-E}{2k_B T}\right)\nonumber\\
&&\times\sum_\tau N_\tau (E){\mbox{Im}}\left\{
T_x^{\tau}(E,\theta,m)\right\},
\label{AmBm}
\end{eqnarray}
together with
\begin{eqnarray}
T_x^\tau(E,\theta,m)=&&\sum_{n=-\infty}^{\infty}
\left[(a_{2n+2m}^{\tau})^* a_{2n}^{\tau}-
(d_{2n+2m}^{\tau})^*d_{2n}^{\tau}\right.\nonumber\\
&&\left.+(b_{2n+2m}^{\tau})^* b_{2n}^{\tau}-
(c_{2n+2m}^{\tau})^*c_{2n}^{\tau}\right].
\label{Teh}
\end{eqnarray}
In Eq.~(\ref{AmBm}) we write $2f(E)-1=\tanh(-E/2k_BT)$.

Finally, we introduce an amplitude
$C_{x,m}=(A_{x,m}^2+B_{x,m}^2)^{1/2}$ and a phase
$\alpha_{x,m}=\arctan(B_{x,m}/A_{x,m})$ in order to write the current
as
\begin{eqnarray}
\frac{I_x}{\sigma_0}&=&\frac{eV}{\Delta_0}+
\sum_mC_{x,m}\cos(\Omega_m-\alpha_{x,m})
\nonumber\\
&=&\frac{eV}{\Delta_0}+\sum_m C_{x,m}
e^{i(\Omega_m-\alpha_{x,m})},
\label{Icos_Ie}
\end{eqnarray}
where we in the last step used the symmetries $A_{x,-m}=A_{x,m}$,
$B_{x,-m}=-B_{x,m}$ giving $C_{x,-m}=C_{x,m}$,
$\alpha_{x,-m}=-\alpha_{x,m}$.  These symmetries are due to the
relation $T^{\tau}(E,\theta,-m)=[T^{\tau}(E,\theta,m)]^*$. Note that
the last expression in Eq.~(\ref{Icos_Ie}) is equal to
Eq.~(\ref{ac_curr_x}).

Studying the phases $\Omega_m-\alpha_{x,m}$ in Eq.~(\ref{Icos_Ie}) we
see that for $m=0$ they are all zero implying that the zero-frequency
component does not depend on the phase difference $\phi_0$ over the
junction.  For the components with $m\neq 0$, we see that we can
arrange the phases as $\Omega_m-\alpha_{x,m}=m\omega_J (t-t_0)$, where
$t_0=(\alpha_{x,m}/m-\phi_0)/\omega_J$ can be thought of as the time
when the measurement starts. It will therefore be of no importance for
the different current components in the following discussion, which
rather focuses on the amplitude $C_m$.

\subsection{The $y$-current}

We now move on to study the $y$-current (current parallel to the
junction). Below we will see that in general the $y$-current density
$j_y(x)$ (current per $ab$-plane and unit length in the $x$-direction)
depends on the $x$-direction. To stress this dependence on $x$ we have
chosen to discuss the parallel current in terms of the current density
$j_y$.

The analysis of the parallel current density follows closely the one
presented for the perpendicular current above. There are four main
differences: first, the derivative with respect to $y$ in
Eq.~(\ref{currentdensity}) now leads to a factor of $\sin \theta$
(instead of $\cos \theta$); second, there will only be plus signs in
the equation corresponding to Eq.~(\ref{Teh}); third, there is an
oscillating term in $j_y$; fourth, there is no normal contribution to
the parallel current as in the perpendicular case discussed above.

Therefore, we write down (suppressing the $x$-dependence) the parallel
current density $j_y$ as
\begin{equation}
j_y(V,t)=\sum_m j_{y,m}(V) e^{im\omega_J t}
\label{y-FourCurr}
\end{equation}
with the components defined as
\begin{eqnarray}
&&\frac{j_{y,0}(V)}{(\sigma_0/L_y)}=A_{y,0}(V),\nonumber\\
&&\frac{j_{y,m}(V)}{(\sigma_0/L_y)}=C_{y,m}(V)
e^{i(m\phi_0-\alpha_{y,m})},\ \ m\neq0 \nonumber\\
&&C_{y,m}(V)=\sqrt{A^2_{y,m}(V)+B^2_{y,m}(V)},\nonumber\\
&&\alpha_{y,m}=\arctan\frac{B_{y,m}(V)}{A_{y,m}(V)},
\label{y-FourComponents}
\end{eqnarray}
together with
\begin{eqnarray}
A_{y,m}(V)=&&\frac{1}{4D}\int_{-\pi/2}^{\pi/2}d\theta \sin\theta\
\int_{-\infty}^{\infty}\frac{dE}{\Delta_0}
\tanh\left(\frac{-E}{2k_B T}\right)\nonumber\\
&&\times\sum_\tau \tau N_\tau (E){\mbox{Re}}\left\{
T^{\tau}_y(E,\theta,m)\right\},\nonumber\\
B_{y,m}(V)=&&-\frac{1}{4D}\int_{-\pi/2}^{\pi/2}d\theta \sin\theta
\int_{-\infty}^{\infty}\frac{dE}{\Delta_0}
\tanh\left(\frac{-E}{2k_B T}\right)\nonumber\\
&&\times\sum_\tau N_\tau (E){\mbox{Im}}\left\{
T^{\tau}_y(E,\theta,m)\right\},\nonumber\\
T^\tau_y(E,\theta,m)=&&\sum_{n=-\infty}^{\infty}\left\{
(a_{2n+2m}^{\tau})^* a_{2n}^{\tau}+
(b_{2n+2m}^{\tau})^*b_{2n}^{\tau}\right.\nonumber\\
&&+(c_{2n+2m}^{\tau})^* c_{2n}^{\tau}+
(d_{2n+2m}^{\tau})^*d_{2n}^{\tau}\nonumber\\
&&+2\left[(a_{L,n+m}^{\tau})^*d_{L,n}^{\tau}+
(b_{L,n+m}^{\tau})^*c_{L,n}^{\tau}\right]
\nonumber\\
&&\left.\times e^{-2ik_F\cos\theta x}\right\}.
\label{y-AmBm}
\end{eqnarray}
The last term in $T_y^{\tau}(E,\theta,m)$ oscillates on a scale
$1/k_F$ and would be hard to observe in an experiment. We therefore
drop it when we calculate the parallel current. This is reasonable
since $j_y$ is a current density: To get the total current we should
integrate in the $x$-direction. The oscillating term will then average
to zero.  Dropping the oscillating term, $j_y$ will no longer be
continuous when passing the barrier: the value of $j_y$ will be
different at $x=0^+$ and $x=0^-$ (with the $\delta$-function barrier
located at $x=0$). However, we have checked that the $m=0$ component
of $j_y$ is continuous if the oscillating term is included. For
completeness we will in this paper give (the average value of) the
current density on both sides of the barrier, denoted $j_{yL}$ and
$j_{yR}$, respectively. For the ballistic case ($Z=0$) we have
$j_{yL}=j_{yR}$.

In the next section we discuss $C_{x,m}(V)$ and $C_{y,m}(V)$ for some
configurations of superconductors.

\section{Results and discussion}\label{Sec_4}
We will in this section present curves only for positive voltage: in
Appendix~\ref{App_C} we show that the current is odd in voltage.

\begin{figure}
\centerline{\psfig{figure=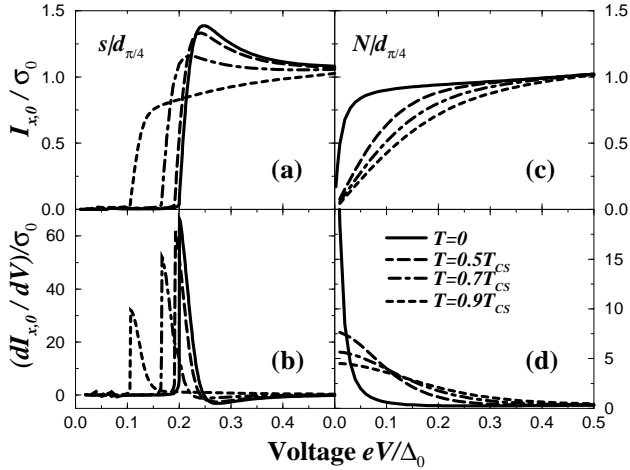,width=8cm}}
\caption{Current-voltage curves and differential
  conductance-voltage curves for $s|d_{\pi/4}$ [(a) and (b)
  respectively] and $N|d_{\pi/4}$ junctions [(c) and (d) respectively]
  are shown for some temperatures. The zero bias conductance peak
  present for the $N|d$ case is moved to finite bias $eV=\Delta_s(T)$
  when the counter electrode is an $s$-wave superconductor. In
  addition negative conductance appears. The peak moves to smaller
  voltage when $T$ is close to $T_{cs}$ since $\Delta_s$ is reduced by
  temperature according to the BCS theory. The figure demonstrates
  that the peak is less sensitive to temperature for the $s|d$
  junction compared to the $N|d$ junction. The barrier strength is
  Z=5, and $\Delta_s(T=0)=0.2\Delta_0$.}
\label{fig5}
\end{figure}

We start by discussing the results for the $m=0$ term of the sum in
Eq.~(\ref{ac_curr_x}). In Fig.~\ref{fig5} we have plotted the $I-V$
curves for different values of the temperature. For zero temperature
there is a peak in the current for some orientations due to resonant
transport through MGS.\cite{Hurd,BarSvi1,BarSvi2} This feature
(resulting in negative differential conductance) is especially
pronounced for arrangements where one superconductor is an $s$-wave
superconductor and the other one is a $d$-wave superconductor with
orientation angle $\alpha=\pi/4$. In Fig.~\ref{fig5}(a)-(b) we study
how temperature influences the current peak appearing at a voltage
$eV=\Delta_s$. We find that the peak is rather robust up to relatively
high temperatures, such as $0.5T_{cs}$ shown in the figure, where
$T_{cs}$ is the critical temperature of the $s$-wave superconductor.
In Fig.~\ref{fig5} we have taken into account the fact that the
$s$-wave superconducting gap is reduced when the temperature is
increased according to the theory by Bardeen, Cooper, and
Schrieffer.\cite{BCS} This is the reason for the shift of the peak
towards lower voltage when the temperature is increased. The $d$-wave
order parameter, on the other hand, is assumed not to be affected by
the temperature variations, since $T_c$ for HTS compounds in general
is considerably higher compared to conventional superconductors.

It is interesting to compare with the NS junction,\cite{TanKas} where
it has been found that the presence of MGS induces a dramatic increase
of current for low voltage resulting in ZBCP. This structure is more
sensitive to temperature than the current peak of the
$s|d_{\pi/4}$-junction discussed in the paragraph above. The reason
for this greater sensitivity is the density of states: it is well
known that MGS are located at the Fermi energy on the $d$-wave side.
This means that quasiparticles incident from the normal side will
reach MGS for any small voltage. This, however, is in great contrast
to the $s|d_{\pi/4}$-case, where MGS on the $d$-wave side may only be
reached from the $s$-wave side at a voltage greater than $\Delta_s$
simply because the superconducting density of states is zero for
subgap energies. The conclusion is therefore that the NS case will be
more affected by thermal smearing introduced by the Fermi functions,
since this smearing is centralized around the Fermi energy. This
argument breaks down at temperatures larger than the $s$-wave gap,
taking place close to $T_c$ of the $s$-wave superconductor.

An interesting paper\cite{SinNg} has recently appeared studying
experimentally the system corresponding to the case depicted in
Fig.~\ref{fig5}.  In that paper two sets of curves are presented:
first, Fig.~4(a) of Ref.~\onlinecite{SinNg} denoted here by (i);
second, Fig.~4(b) of Ref.~\onlinecite{SinNg} denoted here by (ii).
While the results of (i) are in agreement with the picture outlined in
our Fig.~\ref{fig5}, the curves of (ii) are not easily understood from
our calculations. The critical point for not having agreement is the
fact that in (ii) there is no opening of the superconducting $s$-wave
gap corresponding to the Pb electrode to be found.

\begin{figure}
\centerline{\psfig{figure=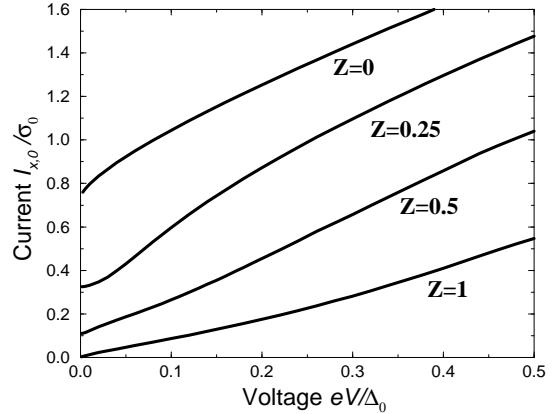,width=8cm}}
\caption{Current-voltage curves are plotted for the
  $d_{\pi/8}|d_{-\pi/8}$ junction with barrier strenths $Z=0$, $0.25$,
  $0.5$, and $1$.  The finite value of the current at small voltage
  for low (but finite) barrier strength is due to resonant transport
  through MGS making the junction ballistic. For comparison, the
  ballistic ($Z=0$) curve is shown (no MGS in the ballisic case) where
  there is always a finite current in the zero-bias limit.}
\label{fig6}
\end{figure}

Very recently an investigation of ZBCP in bicrystal grain boundary
junctions have appeared,\cite{Alff2,Alff3} discussing (among other
things) a geometry corresponding to $\alpha_1=-\alpha_2=18.4^{\circ}$,
in which ZBCP was found. In our calculations ZBCP can only be found in
rather transparent (small $Z$) $d|d$ junctions, as shown in
Fig~\ref{fig6}. The reason for this is that the scattering states
start from the bulk where there are no superconducting density of
states close to zero energy except for angles corresponding to the gap
nodes of the $d$-wave superconductor.  For small voltage, the MGS can
only be reached by the scattering states from those angles, making the
MGS contribution very limited in the angle average. Broadening of MGS
due to reduced barrier height $Z$ (which is incorporated in our
treatment from the start) increases the range of angles from which the
scattering states may reach MGS. For low barrier height the width of
the MGS becomes comparable to the gap making the junction
ballistic\cite{GunZai,HDB1} for scattering states starting near the
gap edges, allowing MAR to contribute massively to the current.  The
current then takes a finite value for small voltage and a huge ZBCP
appears. It is important to have MGS on both sides of the barrier for
this effect; otherwise, the number of AR will be reduced, resulting in
a smaller current contribution. Fig.~\ref{fig6} demonstrates that at a
barrier strength $Z=1$, corresponding to an averaged transparancy
$D=0.38$, the ZBCP has disappeared within our model: for this barrier
strength the width of the MGS is not high enough to give large current
at small voltage.

A related issue is the case $\alpha_1=\alpha_2=45^{\circ}$ discussed
in Ref.~\onlinecite{Hurd}. Also in this case a combination of
broadened MGS on both sides of the barrier and MAR give rise to ZBCP
for small (e. g. $Z=0.25$) barrier strengths, although the peak is not
as sharp as for $d_{\pi/8}|d_{-\pi/8}$. As explained above there are
no scattering states at zero energy except for the angles
corresponding to the gap nodes. Since for the orientation
$\alpha_1=\alpha_2=45^{\circ}$ the gap nodes are geometrically lined
up against each other, the influence of MGS will be less dramatic
compared to the case discussed in the previous paragraph. An
interesting paper, mainly on calculations in the tunneling limit,
introduces phenomenologically an imaginary part of the energy,
corresponding to broadening of the MGS.\cite{SamDat} (In our
calculation there is no imaginary part of the energy.) Due to the
imaginary part ZBCP was found in the case
$\alpha_1=\alpha_2=45^{\circ}$.\cite{SamDat} The difference compared
to our work is that in our work the broadening is not introduced
phenomenologically; contrary, it is microscopically described in terms
of reduced strength of the barrier.

\begin{figure}
\centerline{\psfig{figure=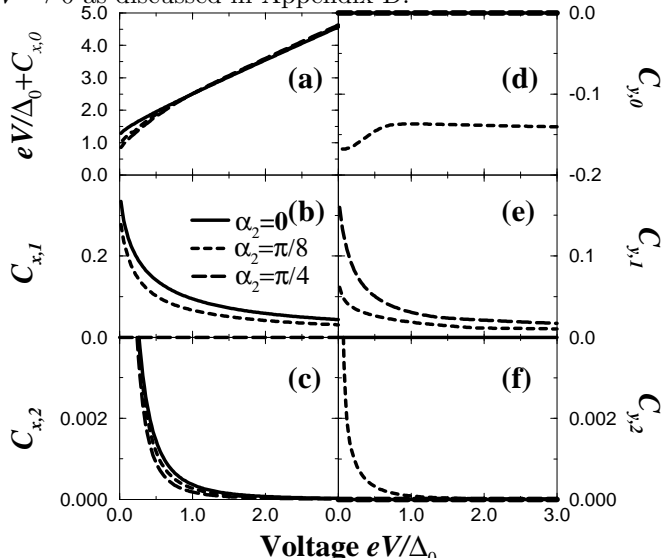,width=9cm}}
\caption{The first three components ($m=0$, $1$, $2$) of the
  perpendicular current [(a)-(c)] and the parallel current [(d)-(e)]
  is shown for the $d_0|d_{\alpha_2}$ junction for three different
  orientations of the right superconductor. The odd components of the
  perpendicular current is suppressed when the orientation angle
  $\alpha_2$ is changed from $0$ towards $\pi/4$ (and zero at
  $\pi/4$). At the same time the odd component of the parallel current
  is enhanced. We also see that all components of the parallel current
  vanishes for the orientation $\alpha_2=0$. The barrier strength is
  $Z=0$ and the temperature is zero.}
\label{fig7}
\end{figure}

We now continue to discuss the $m\neq 0$ ac current components of
Eq.~(\ref{ac_curr_x}). In Fig.~\ref{fig7} we show some results for
current density for both the perpendicular ($x$) and parallel ($y$)
direction. The curves shown in Fig.~\ref{fig7} are for the ballistic
($Z=0$) case. The ballistic case is rather restricted; however, it
will prove very useful for us to illustrate some important properties
of $d_{\alpha_1=0}|d_{\alpha_2}$ junctions in the ac case.  These
qualitative properties carry over to the non-ballistic case in a
straightforward manner. Quantitatively, the curves will look different
in the presence of normal scattering.

We start by discussing $\alpha_2 =\pi/4$.  In Fig.~\ref{fig7} we plot
the magnitude $C_m$ of the ac components. It is then possible for us
to show how the disappearance of the ac components in the
$x$-direction\cite{LJHW} will be complemented by the appearance of the
corresponding ac components in the $y$-direction as seen in
Fig.~\ref{fig7}(b) and (e). The detailed explanation for this effect
is given in Appendix~\ref{App_C} and is good for any value of $Z$. The
main point is that the current is calculated from an integration over
angles $\theta$, where the integrand is composed of two functions. The
first function is $\cos(\theta)$ [$\sin(\theta)$] produced by a
derivative in $x$ ($y$) in Eq.~(\ref{currentdensity}). The other
function can be proved to be odd (even) in $\theta$ for odd (even)
$m$. It is now clear that the combination of these two functions
integrated will result in zero current density in the $x$ ($y$)
direction for odd (even) $m$.  This scenario is confirmed in detail
from the calculation presented in Fig.~\ref{fig7}. We note in passing
that the $m=0$ component has a finite current in the limit
$eV\rightarrow 0$ as discussed in Appendix~\ref{App_D}.

\begin{figure}
\centerline{\psfig{figure=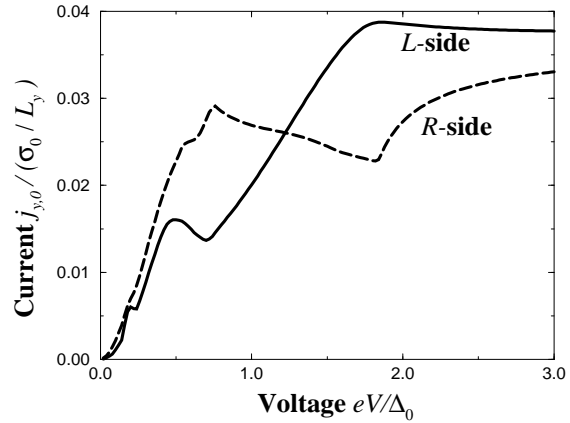,width=8cm}}
\caption{Parallel current for the $d_{0}|d_{\pi/8}$ junction to
  the left and to the right of the barrier. The barrier strength is
  Z=1 and the temperature is zero.}
\label{fig8}
\end{figure}

Next we discuss the parallel current density in the presence of normal
scattering. We specialize to the $d_0|d_{\pi/8}$ junction with $Z=1$
for zero temperature. The curves depicted in Fig.~\ref{fig8} are the
$m=0$ component calculated in the normal regions. As explained in
Section~\ref{Sec_3} the average current density in the left side
differs from the right side of the barrier. We note that the
large-voltage limit of the curves in Fig.~\ref{fig8} is of the order
of the perpendicular excess current density, implying a saturation at
large voltages. Below we indicate which side the current densities are
to be associated with by adding an index $L/R$ for left/right whenever
the side of the junction matters.

The fine structure in Fig.~\ref{fig8} is due to subharmonic gap
structure of the same origin as in the case of perpendicular
current.\cite{Hurd,LJHW}

\begin{figure}
\centerline{\psfig{figure=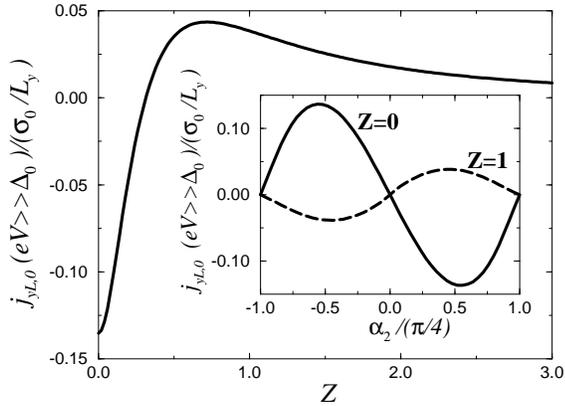,width=8cm}}
\caption{We show the current density parallel to the junction
  interface in the limit $eV>>\Delta_0$ as a function of the barrier
  strength for the orientation $\alpha_1=0$ and $\alpha_2=\pi/8$.  In
  the inset we show the dependence on the misorientation angle
  $\alpha_2$ of the right superconductor for the barrier strengths
  $Z=0$ and $Z=1$ keeping $\alpha_1=0$. Note that the factor $D$
  (which is $Z$-dependent) is included in $\sigma_0$, giving a larger
  normalized value of the current for non-zero $Z$. Zero temperature
  is assumed in all cases.}
\label{fig9}
\end{figure}

The large-voltage limit of the parallel current density can be
calculated analytically in the ballistic limit, see
Appendix~\ref{App_D}. For the orientation $\alpha_1=0$ and
$\alpha_2=\pi/8$, the value of the large-voltage limit of the parallel
current density will be negative. In Fig.~\ref{fig9} we plot the
dependence of the large-voltage limit of the parallel current density
on the barrier strength $Z$. The result is that we get a sign change
at a rather small value of $Z$.  The sign (change) can be
qualitatively understood by studying Fig.~\ref{fig3} in combination
with Eq.~(\ref{largeV_currents}). We see that Fig.~\ref{fig3}
approximately represent the case $\alpha_2=\pi/8$. For $Z=0$ the
current contributions for each angle of incidence will be averaged by
the function $\sin\theta$. We see in Fig.~\ref{fig3} that the largest
gap appears at negative angles where $\sin\theta$ is negative.  The
value of the integral in Eq.~(\ref{largeV_currents}) will therefore be
negative, since only the absolute value of the gap appears in
Eq.~(\ref{largeV_currents}). Introducing normal scattering means that
small angles of incidence are more strongly weighted (note the
structure of transmission amplitude $t(\theta)$). In Fig.~\ref{fig3}
we see that the largest gap will be at positive angles in this case,
implying a positive current since $\sin\theta$ is positive for
positive angles.  In addition, the normal scattering will introduce
contributions from the gap $\bar{\Delta}_2=\Delta_2(-\alpha_2)$ at the
expense of $\Delta_2$. One can show that for the $d_0|d_{\alpha_2}$
junction the relation $j_y(-\alpha_2)=-j_y(\alpha_2)$ holds for all
$Z$, implying that the introduction of $\bar{\Delta}_2$ gives
contributions with the positive sign. The inset in Fig.~\ref{fig9}
illustrates the relation $j_y(-\alpha_2)=-j_y(\alpha_2)$. In
conclusion, the positive contributions will quickly dominate over
negative contributions in the angle average when we change $Z$ from
zero to finite values.

\begin{figure}
\centerline{\psfig{figure=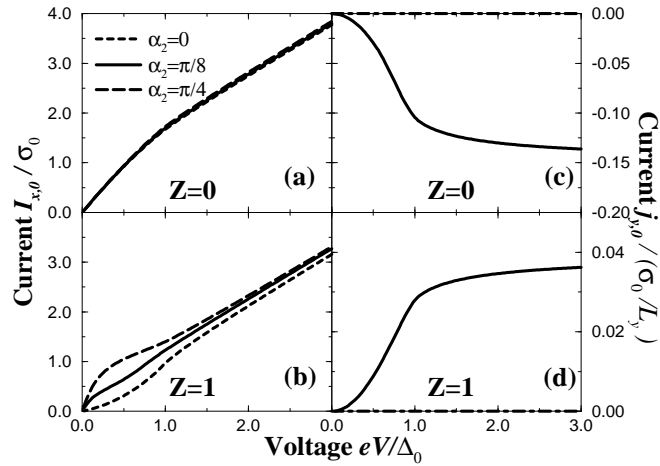,width=8cm}}
\caption{Currents for the $N|d_\alpha$ junction at barrier
  strengths $Z=0$ and $Z=1$. In (a)-(b) we show the perpendicular
  current $I_x$ flowing through the junction and in (c)-(d) we show
  the parallel current density $j_y$. In (b) we see that the zero bias
  conductance peak grows up when we rotate the orientation of the
  superconductor from $\alpha_2=0$ to $\alpha_2=\pi/4$. The parallel
  current only exists when $\alpha_2\neq n\pi/4$, as seen in (c)-(d).}
\label{fig10}
\end{figure}

Finally, we show in Fig.~\ref{fig10} some $I-V$ curves for the
$N|d_{\alpha}$ junction for several barrier strengths and orientations
$\alpha$. The curves demonstrate what is illustrated in
Fig.~\ref{fig3}: for almost any orientation (the exceptions are
$\alpha_2=n\pi/4$) there is a net parallel current density. In our
model the parallel current density will be present everywhere in the
normal region. This might not be surprising since we do not take into
account any impurity scattering in the bulk.

An important limitation of the present work is that we do not include
magnetic field. From a calculation with a self-consistent magnetic
field one should expect that the parallel current density is screened
out in the superconductors on the length scale of the London
penetration depth.  This means that our results for the parallel
current density presented in Figs.~\ref{fig7} to \ref{fig10} are only
valid for systems with dimensions smaller than the London penetration
depth.

\section{Summary}\label{Sec_5}
In this paper we have studied the ac Josephson effect for planar
superconducting $d$-wave junctions. The presence of zero-energy midgap
states (MGS) at interfaces of $d$-wave superconductors for some
orientations influence the current-voltage curves (the zeroth order ac
component), since MGS open up additional channels for current flow. We
have primarily discussed two aspects of MGS in this paper. First, our
results feature current peaks at finite voltage (resulting in negative
differential conductance) in the $\mbox{S}_1|\mbox{S}_2$
case.\cite{Hurd} We have in this paper explored how sensitive this
current peak is to temperature. We have also discussed how well our
calculations compare with an experiment:\cite{SinNg} some curves of
this experiment compare well with our results; however, there is in
Ref.~\onlinecite{SinNg} a set of curves which we find difficult to
understand within the framework of the model we are using. The second
aspect relates to a recent experiment on $d_{\alpha}|d_{-\alpha}$
junctions,\cite{Alff2,Alff3} where zero-bias conductance peaks
(corresponding to large current increase for small valtage) have been
found. Our model explains these experimental results in terms of
resonant transport through MGS if the grain boundary is rather
transparent, allowing multiple Andreev reflections to contribute
hugely to the current for low voltage.

An interesting feature of superconducting $d$-wave junctions is the
possiblity to have parallel current density at the interface of the
two electrodes. We explain the reason for this effect for the ac
Josephson effect.  We also demonstrate how the disappearance of the
odd ac components for the perpendicular current\cite{LJHW} in some
orientations is complemented by appearance of odd ac components in the
direction parallel to the junction.

Moreover, we have given a fairly complete account of our method which
is applicable to all types of junctions: all the way from the
ballistic (no barrier) to the insulating case. We have also discussed
how different current formulas previously used relate to each other.

\section*{Acknowledgments}
It is a pleasure to thank V. S. Shumeiko and M. Moraghebi for many
useful discussions on this work. M. H. also acknowledges discussions
with P. Johansson. Finally, we thank L. Alff for explaining some
geometrical aspects of the grain boundary experiments performed in
Ref.~\onlinecite{Alff2}.

\appendix

\section{Solution of the Bogoliubov-de Gennes equation}\label{App_A}
The complicated integral operator makes exact solutions of
Eq.~(\ref{BdG}) not easily accessible. Therefore, we solve
Eq.~(\ref{BdG}) approximately by an ansatz in each region of the
following type:
\begin{eqnarray} 
\left\lgroup
\begin{array}{c}
u({\bf x},t) \\
v({\bf x},t)
\end{array}
\right\rgroup
=
\sum_{\bf k}
A_{\bf k}  e^{i{\bf k}\cdot {\bf x}}
\left\lgroup
\begin{array}{c}
u_{\bf k} \\
v_{\bf k}
\end{array}
\right\rgroup
e^{-iE_{\bf k}t/\hbar}
\; ,
\label{general_ansatz}
\end{eqnarray}
where the Fourier sum over different frequencies reflects the fact
that there is a time dependence in the order parameter after a global
reference level is chosen as described in Section~\ref{Sec_2}. We
follow the standard treatment when approximating the integral operator
of Eq.~(\ref{BdG}): the center of mass coordinate ${\bf R}=({\bf
  x}+{\bf x}')/2$ and the relative mass coordinate ${\bf r}={\bf
  x}-{\bf x}'$ are introduced. This means that in these new
coordinates the order parameter is written $\Delta({\bf x},{\bf
  x}')=\tilde{\Delta}({\bf r},{\bf R})$. The integral operator of
Eq.~(\ref{BdG}) is now approximated as
\begin{eqnarray}
&&\int d{\bf x}' \Delta({\bf x},{\bf x}')v({\bf x}',t) =
\int d{\bf r} \tilde{\Delta}({\bf r},{\bf x}-{\bf r}/2)
v({\bf x}-{\bf r},t) \nonumber \\
&&\approx\sum_{\bf k} \Delta({\bf k},{\bf x})A_{\bf k}v_{\bf k}
e^{i{\bf k}\cdot{\bf x}}e^{-iE_{\bf k}t/\hbar}\;,
\label{approx}
\end{eqnarray}
where we have introduced the Fourier transform $\Delta({\bf k},{\bf
  R})$ of the order parameter $\tilde{\Delta}({\bf r},{\bf R})$.  In
Eq.~(\ref{approx}) we approximated $\tilde{\Delta}({\bf r},{\bf
  x}-{\bf r}/2)\approx \tilde{\Delta}({\bf r},{\bf x})$, valid since
$\tilde{\Delta}({\bf r},{\bf x})$ is assumed to be a slow (fast)
function in ${\bf x}$ (${\bf r}$). Since in the weak-coupling limit
the pair potential is expected to be non-zero only close to $k_F$, one
can replace the momentum by an angle $\theta$. The approximation made
in Eq.~(\ref{approx}) neglects terms of the order $(k_F\xi_0)^{-1}$
and is called the quasi-classical approximation.\cite{Brud}
  
Before solving the time-dependent BdG equation it is convenient to
introduce some notational simplifications.  First we define the BCS
coherence factors
\begin{eqnarray}
&&\frac{v_i(E,\theta)}{u_i(E,\theta)}=
\frac{E-\mbox{sgn}(E)\sqrt{E^2-\Delta_i(\theta)^2}}{\Delta_i(\theta)},\;
|E|>|\Delta_i(\theta)| \nonumber \\
&&\frac{v_i(E,\theta)}{u_i(E,\theta)}=
\frac{E-i\sqrt{\Delta_i(\theta)^2-E^2}}{\Delta_i(\theta)},\;
|E|<|\Delta_i(\theta)| 
\label{uv}
\end{eqnarray}
where $|u_i|^2+|v_i|^2=1$ and $i=1,2$ refers to superconductor 1 and
2. Using this definition, one can express the Andreev reflection
amplitude at superconductor $i$ as $A_i(E_n,\theta)\equiv
A_{i,n}=v_i(E_n,\theta)/u_i(E_n,\theta)$, where $E_n=E+neV$.  The
transmission amplitude for an electron-like quasiparticle at energy
$E$ and angle $\theta$ from superconductor $i$ to enter the electron
branch in the normal region is $J_{i}(E,\theta)\equiv
J_i=(u_{i}^2(E,\theta)-v_{i}^2(E,\theta))/u_{i}(E,\theta)$. An
important notational detail in this paper are the barred quantities:
these are to be associated with the angle $\bar{\theta}=\pi-\theta$.

To solve the time-dependent BdG equation we follow the method worked
out for the $s$-wave case\cite{BSW,AveBar,HDB1,HDB2} and further
generalized to the anisotropic case.\cite{Hurd,LJHW} We use the
boundary condition that far away from the junction the wave function
corresponds to an electron-like quasiparticle incident on the junction
from one side at an energy $E$ and an angle $\theta$. Thanks to
symmetries of the BdG equation it is enough to study these
electron-like quasiparticle solutions as explained in
Appendix~\ref{App_B}. The incident quasiparticle gives in general rise
to both reflected and transmitted waves.  Therefore, we start writing
down the wave function in the left superconductor as (corresponding to
a quasiparticle incident on the junction from the left)
\begin{eqnarray}
{\Psi}_{S_{1}}^{\rightarrow}
&=&\sum_n\left[\delta_{n,0}\left(
\begin{array}{c}
u_{1}(E_n,\theta)\\
v_{1}(E_n,\theta)
\end{array}
\right)\right.
e^{i {\bf q}^{e}\cdot{\bf x}}
\nonumber\\
&+&b_{1,n}
\left(
\begin{array}{c}
v_{1}(E_n,\theta)\\
 u_{1}(E_n,\theta)
\end{array}
\right)
e^{i {\bf q}^{h}\cdot{\bf x}}
\nonumber\\
&+&\left. d_{1,n}
\left(
\begin{array}{c}
\bar{u}_{1}(E_n,\theta)\\
\bar{v}_{1}(E_n,\theta)
\end{array}
\right)
e^{i {\bf\bar{q}}^{e}\cdot{\bf x}}\right]
e^{-i (\frac{E_n t}{\hbar}+\frac{n\phi_0}{2})}.
\label{wfnS1}
\end{eqnarray}
where ${\bf q}^{e/h}$ corresponds to the wave vector of
electron/hole-like quasiparticles.  For the wave vectors above (and in
the following) we use the notation ${\bf
  q}=(q_x,q_y)=q(\cos\theta,\sin\theta)$ and $\bar{\bf
  q}=(-q_x,q_y)=q(\cos\bar{\theta},\sin\bar{\theta})$, where ${\bar
  \theta}=\pi-\theta$ corresponds to the angle after normal
scattering.  We will approximate the magnitude of any wave vector in
this paper as the Fermi wave vector, $q \approx k_F$.  Again, all
barred quantities are to be associated with the angle ${\bar \theta}$.
In the voltage case the injected quasiparticle goes through multiple
Andreev reflections (MAR). The wavefunction therefore contains
sidebands at energies $E_n=E+neV$, where $n$ is an even integer.

In the normal region the wave function to the left and right of the
barrier is
\begin{eqnarray}
&&{\Psi}_{L}^{\rightarrow} =
\sum_{n} 
\left(
\begin{array}{c}
a_{L,n}e^{i {\bf k}^e\cdot{\bf x}}
+d_{L,n} e^{i {\bf\bar{k}}^e\cdot{\bf x}}\\
b_{L,n} e^{i {\bf k}^h\cdot{\bf x}} 
+c_{L,n} e^{i {\bf \bar{k}}^h\cdot{\bf x}} 
\end{array}
\right)
e^{-i(\frac{E_n t}{\hbar}+\frac{n\phi_0}{2})},\\
&&{\Psi}_{R}^{\rightarrow} =
\sum_{n} 
\left(
\begin{array}{c}
a_{R,n}e^{i {\bf k}^e\cdot{\bf x}}
+d_{R,n} e^{i {\bf\bar{k}}^e\cdot{\bf x}}\\
b_{R,n} e^{i {\bf k}^h\cdot{\bf x}} 
+c_{R,n} e^{i {\bf \bar{k}}^h\cdot{\bf x}} 
\end{array}
\right)
e^{-i(\frac{E_n t}{\hbar}+\frac{n\phi_0}{2})},
\label{wfnR}
\end{eqnarray}
respectively. The wave function of the right superconductor is
\begin{eqnarray}
{\Psi}_{S_{2}}^{\rightarrow}&=&
\sum_{n}\left\{\left[
a_{2,n}
\left(
\begin{array}{c}
u_{2}(E_n,\theta)e^{i\frac{\phi_0}{2}}e^{i(\frac{eVt}{\hbar})}\\
v_{2}(E_n,\theta)e^{-i\frac{\phi_0}{2}}e^{-i(\frac{eVt}{\hbar})}
\end{array}
\right)
e^{i {\bf p}^{e}\cdot{\bf x}}\right.\right.\nonumber\\
&&+\left. c_{2,n}
\left(
\begin{array}{c}
\bar{v}_{2}(E_n,\theta)e^{i\frac{\phi_0}{2}}e^{i(\frac{eVt}{\hbar
})}\\
\bar{u}_{2}(E_n,\theta)e^{-i\frac{\phi_0}{2}}e^{-i(\frac{eVt}{\hbar})}
\end{array}
\right)
e^{i {\bf\bar{p}}^{h}\cdot{\bf x}}\right]
\nonumber\\
&&\left.\times e^{-i (\frac{E_n
t}{\hbar}+\frac{n\phi_0}{2})}\right\} \; .
\label{wfnS2}
\end{eqnarray}

In the wave functions written down in Eqs.~(\ref{wfnS1})-(\ref{wfnS2})
we have explicitly factored out the dependence on the phase $\phi_0$
for convenience, making the coefficients $a$, $b$, $c$, and $d$ all
independent of $\phi_0$. In the final expression for the current,
however, the phase $\phi_0$ will reappear.

In the following we solve the matching problem. We have three
interfaces where the wave functions and derivatives in
Eqs.~(\ref{wfnS1})-(\ref{wfnS2}) are to be matched.  The procedure is
the same as the one carried out in Refs.~\onlinecite{HDB2} and
\onlinecite{HDB1}.  When matching at the left interface between
superconductor 1 and the left normal region we get
\begin{equation}
a_{L,n}=A_{1,n}b_{L,n}+J_1\delta_{n,0},\ \ \
c_{L,n}=\bar{A}_{1,n}
d_{L,n}\;,
\label{ac_L}
\end{equation}
after eliminating coefficients pertaining to the left superconductor.

Repeating the procedure at the interface between the right normal
region and superconductor 2 we have
\begin{equation}
b_{R,n+2}=A_{2,n+1}a_{R,n},\ \ \ d_{R,n}=\bar{A}_{2,n+1}
c_{R,n+2}\;.
\label{ac_R}
\end{equation}
We have now eliminated coefficients related to the superconducting
regions.

It now remains to match the wave functions in the normal region at the
scattering center. This procedure may be expressed in terms of the
scattering matrix
\begin{equation}
S_e(\theta)=\left(
\begin{array}{cc}
r(\theta) & t(\theta)\\
t(\theta) & -r^*(\theta)t(\theta)/t^*(\theta)
\end{array}
\right).
\label{s_matrix}
\end{equation}
The scattering matrix for holes is $S_h(\theta)=S_e^*(\theta)$.  The
scattering matrix relates incoming waves with outgoing waves.  As
outlined in Ref.~\onlinecite{HDB2} for the case of $s$-wave
superconductors with different gaps, the wave-function coefficients of
the left and right side of the normal region fulfill [suppressing the
$\theta$-dependence of $r(\theta)$ and $t(\theta)$]
\begin{eqnarray}
&&d_{L,n}=rA_{1,n}b_{L,n}+r J_1\delta_{n,0}+
t\bar{A}_{2,n+1}c_{R,n+2},\label{EQdabc1}\\
&&a_{R,n}=tA_{1,n}b_{L,n}+tJ_1\delta_{n,0}-
\frac{r^*t}{t^*}\bar{A}_{2,n+1}c_{R,n+2},\label{EQdabc2}\\
&&b_{L,n}=r^*\bar{A}_{1,n}d_{L,n}+
t^*A_{2,n-1}a_{R,n-2},\label{EQdabc3}\\
&&c_{R,n}=t^*\bar{A}_{1,n}d_{L,n}-
\frac{r t^*}{t}A_{2,n-1}a_{R,n-2}.\label{EQdabc4}
\end{eqnarray}

Using Eqs.~(\ref{ac_L})-(\ref{EQdabc4}) we solve for $d_L$. As a
result we find the following recursive relation:
\begin{equation}
\alpha_n d_{L,n+2}+\beta_n d_{L,n}+\gamma_n
d_{L,n-2}=rJ_1\delta_{n,0},
\label{recd}
\end{equation}
where
\begin{eqnarray}
&&\alpha_n=-|t|^2\frac{\bar{A}_{2,n+1}\bar{A}_{1,n+2}}
{1-A_{2,n+1}\bar{A}_{2,n+1}}
\nonumber \\
&&\beta_n=1-A_{1,n}\bar{A}_{1,n}+|t|^2\left(\frac{A_{1,n}\bar{A}_
{1,n}}
{1-A_{2,n-1}\bar{A}_{2,n-1}}
\right.
\label{abc}\\
&&\left.+\frac{A_{2,n+1}\bar{A}_{2,n+1}}{1-A_{2,n+1}\bar{A}_{2,n+
1}}\right)
\;, \ \ \ \ \ \gamma_n=-|t|^2\frac{A_{2,n-1}A_{1,n}}{1-A_{2,n-1}
\bar{A}_{2,n-1}}.  \nonumber
\end{eqnarray}
The coefficient $b_L$ is determined from
\begin{eqnarray}
b_{L,n+2}&=&\frac{1}{r(1-A_{2,n+1}\bar{A}_{2,n+1})}
\left(|t|^2A_{2,n+1}d_{L,n}\right.
\nonumber\\
&&\left.+\bar{A}_{1,n+2}(|r|^2-A_{2,n+1}\bar{A}_{2,n+1})d_{L,n+2}
\right).
\label{b}
\end{eqnarray}
Eqs.~(\ref{recd})-(\ref{b}) and Eq.~(\ref{ac_L}) above, together with
the boundary condition that the coefficients should go to zero as $n$
goes to plus or minus infinity, determine the wave function completely
in the region to the left of the barrier.\cite{Hurd}

Studying Eqs.~(\ref{abc})-(\ref{b}) we notice some unnecessary
singularities, which we may get rid of by defining
\begin{eqnarray}
d_{L,n}'=d_{L,n}/(r\eta_{n-1}\eta_{n+1}),\label{d-prim}\\
\eta_n=1-A_{2,n}\bar{A}_{2,n}\nonumber,
\end{eqnarray}
and
\begin{eqnarray}
\alpha_n '&=&\eta_{n+1}\eta_{n+3}\alpha_n,
\nonumber\\
\beta_n '&=&\eta_{n-1}\eta_{n+1}\beta_n,
\nonumber\\
\gamma_n '&=&\eta_{n-3}\eta_{n-1}\gamma_n.
\label{abc_prim}
\end{eqnarray}
The coefficient $b_L$ is determined from
\begin{eqnarray}
b_{L,n+2}&=&|t|^2A_{2,n+1}\eta_{n-1}d_{L,n}'\nonumber\\
&&+\bar{A}_{1,n+2}(|r|^2-A_{2,n+1}\bar{A}_{2,n+1})\eta_{n+3}d_{L,
n+2}'.
\label{b_new}
\end{eqnarray}
Using the primed quantities above there is a new recursive relation
\begin{equation}
\alpha_n ' d_{L,n+2}'+\beta_n ' d_{L,n}'+\gamma_n '
d_{L,n-2}'=J_1\delta_{n,0},
\label{recd_prime}
\end{equation}
which is more suitable to handle from a numerical point of view.
Solving for $d_{L}'$ we use Eq.~(\ref{d-prim}) to get the original
wave function coefficient $d_{L}$.

To solve Eq.~(\ref{recd_prime}) we use a method previously introduced
in the $s$-wave case.\cite{BSW} First, we introduce
\begin{equation}
x_{n}=\left\{
\begin{array}{ll}
d_{L,n}'/d_{L,n-2}', & n>0\nonumber\\
d_{L,n}'/d_{L,n+2}', & n<0,
\end{array}
\right.
\end{equation}
leading to
\begin{eqnarray}
\alpha_n ' x_{n+2}+\beta_n '+\gamma_n '\frac{1}{x_n}=0, &
n>0\nonumber\\
\alpha_n '\frac{1}{x_n}+\beta_n '+\gamma_n ' x_{n-2}=0, & n<0,
\label{new_recurr}
\end{eqnarray}
together with 
\begin{equation}
d_{L,0}'(\alpha_0 ' x_2+\beta_0 '+\gamma_0 ' x_{-2})=J_1.
\end{equation}
Rearranging Eq.~(\ref{new_recurr}) we get a continued fraction
expansion
\begin{eqnarray}
x_n=\frac{-\gamma_n '}{\beta_n '+\alpha_n ' x_{n+2}}, & n>0,
\nonumber\\
x_n=\frac{-\alpha_n '}{\beta_n '+\gamma_n ' x_{n-2}}, & n<0.
\end{eqnarray}
The continued fraction expansion can be evaluated by a routine in
Ref.~\onlinecite{NumRec} giving us a good approximation for
$x_{n_{large}}$, or be truncated at a large $n=n_{large}$. It turns
out that the truncation method is often very accurate, especially if
the barrier strength $Z$ is large.  Having calculated $x_{n}$ we get
all $d_{n}'$ from the relations
\begin{equation}
d_{L,n}'=\left\{
\begin{array}{ll}
x_n x_{n-2} ...  x_2 d_{L,0}', & n>0\nonumber\\
x_n x_{n+2} ...  x_{-2} d_{L,0}', & n<0.
\end{array}
\right.
\end{equation}

There is a similar procedure as above to calculate the wave function
coefficients of electronlike quasiparticle incident from the right.
One could also get the wave functions for the leftmovers using the
formulas above if the substitutions $V\rightarrow-V$,
$\phi_0\rightarrow -\phi_0$ and $\Delta_{1(2)} \rightarrow
\Delta_{2(1)}$ are made.

In Eqs.~(\ref{AmBm}) and (\ref{y-AmBm}), the density of states in the
superconductors $N_{\tau}(E)$ is included which will give rise to
discontinuities at the gap edges. This will produce numerical errors
when we perform numerical calculations.  Fortunately, the transmission
amplitude $J_{\tau}$ appearing in the recurrence equation
[Eq.~(\ref{recd})] will only enter as a prefactor in the coefficient
$d$, and therefore also in $a$, $b$, and $c$. Hence, a factor
$J_{\tau}^2$ can be extracted from $T^{\tau}(E,\theta,m)$ in
Eq.~(\ref{Teh}) and Eq.~(\ref{y-AmBm}) and combined with the density
of states:
\begin{eqnarray}
N_\tau(E)J_\tau^2(E)&=&
\frac{\theta(|E|-|\Delta_\tau|)}{|u_\tau^2-v_\tau^2|}
\left(\frac{u_\tau^2-v_\tau^2}{u_\tau}\right)^2\nonumber\\
&=&\theta(|E|-|\Delta_\tau|)(1-A_{\tau,0}^2).
\end{eqnarray}
The discontinuities at the gap edges are then eliminated.

Another numerical difficulty is the appearance of a number of
resonances in the continued fraction expansion solution of the
recurrence relation. For a particle coming in at energy $E$ and angle
$\theta$ these resonances appear if the trajectory (MAR state) hits
zero energy (MGS) or the gap edges. When the integration over energy
is performed, it is therefore a good idea to take special care of
these points in order to faster get an accurate result.

\section{Equivalence of current formulas}\label{App_B}
In the literature different current formulas have been used. In this
Appendix we show analytically that they are all the same. In addition
we have noticed that the different formulas numerically give the same
results.

A very useful relation is the completeness relation
\begin{equation}
\sum_{\nu}\left[u_{\nu}({\bf r})u_{\nu}^*({\bf r}')+
v_{\nu}^*({\bf r})v_{\nu}({\bf r}')\right]=\delta({\bf r}-{\bf
r}'),
\label{completeness}
\end{equation}
where the sum runs over solutions to the BdG equation with positive
energy.~\cite{Bardeen} We apply $\nabla_{{\bf r}'}$ and take the
imaginary part:
\begin{equation}
\sum_{\nu}\mbox{Im}\left\{v_{\nu}\nabla v_{\nu}^*\right\}
=-\sum_{\nu}\mbox{Im}\left\{u_{\nu}^*\nabla u_{\nu}\right\}.
\label{uv-relation}
\end{equation}
From now on we let the sum over states (here $\nu$) run over
$k$-vectors, since we in this paper discuss plane wave solutions.

Next, we have the following symmetry between solutions to the BdG
equation with positive and negative energies
\begin{eqnarray}
u_k(E)&=&-v_{-k}^*(-E)\nonumber\\
v_k(E)&=&u_{-k}^*(-E),
\label{BdGsym}
\end{eqnarray}
where the left hand side is a solution describing an electron-like
quasiparticle with energy $E$, and the right hand side is a solution
describing a hole-like quasiparticle with energy $-E$. Both particles
have positive group velocities. We can use Eq.~(\ref{BdGsym}) in
Eq.~(\ref{uv-relation}) to get the same type of completeness also for
negative energies.

\begin{figure}[t]
\psfig{figure=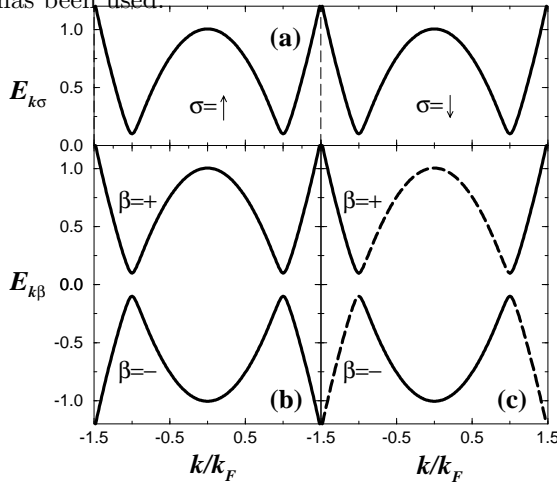,width=8cm}
\caption{Here we show the different sets of states to sum over
  for the different current formulas. In (a) we show the two branches
  with positive energies (labeled by a spin index $\sigma$) included
  in Eq.~(\ref{currentop_exc-pic}). In (b) we show the negative
  ($\beta=-$) and the positive ($\beta=+$) branches included in
  Eq.~(\ref{currentop_semicond-pic}). In (c) we show the states
  included in Eq.~(\ref{currentop_semi-e}): only the electron-like
  particles are included (the solid line), while the hole-like
  particles are excluded (the dashed line).}
\label{fig11}
\end{figure}

The current formula can be written in the excitation
picture~\cite{deG} (positive energies only) as
\begin{eqnarray}
j&=&\frac{e\hbar}{m}\sum_{\sigma}\sum_k\mbox{Im}
\left\{u_{k\sigma}^*\nabla u_{k\sigma}-
v_{k\sigma}\nabla v_{k\sigma}^*\right\}f(E_{k\sigma})\nonumber\\
&+&\frac{e\hbar}{m}\sum_{\sigma}\sum_k\mbox{Im}
\left\{v_{k\sigma}\nabla v_{k\sigma}^*\right\},
\label{currentop_exc-pic}
\end{eqnarray}
where $\sigma=\uparrow$, $\downarrow$ labels spin up and spin down,
see Fig.~\ref{fig11}(a).  Since in this paper a spin independent
problem is featured, the spin sum just produces a factor of two.  One
of the terms (say $\sigma=\downarrow$) in the spin sum can be mapped
into negative energies using Eq.~(\ref{BdGsym}) and the relation
$f(-E)=1-f(E)$. After this mapping to negative energy states the spin
sum transforms into a summation over branches of negative ($\beta=-$)
and positive ($\beta=+$) energies (sometimes denoted the semiconductor
picture), see Fig.~\ref{fig11}(b). As a result the current formula
after these modifications is
\begin{eqnarray}
j&=&\frac{e\hbar}{m}\sum_{\beta}\sum_k\mbox{Im}
\left\{u_{k\beta}^*\nabla u_{k\beta}-
v_{k\beta}\nabla v_{k\beta}^*\right\}f(E_{k\beta})\nonumber\\
&+&\frac{e\hbar}{m}\sum_{\beta}\sum_k\mbox{Im}
\left\{v_{k\beta}\nabla v_{k\beta}^*\right\}.
\label{currentop_semi-pic}
\end{eqnarray}

We now study the term in Eq.~(\ref{currentop_semi-pic}) without the
Fermi-factor.  We write out the sum over $\beta$:
\begin{eqnarray}
&&\sum_{\beta}\sum_k\mbox{Im}
\left\{v_{k\beta}\nabla v_{k\beta}^*\right\}\nonumber\\
&=&\sum_k\left[\mbox{Im}\left\{v_{k+}\nabla v_{k+}^*\right\}
-\mbox{Im}\left\{u_{k-}^*\nabla u_{k-}\right\}\right]\nonumber\\
&=&\sum_k\left[\mbox{Im}\left\{v_{k+}\nabla v_{k+}^*\right\}
-\mbox{Im}\left\{v_{-k,+}\nabla v_{-k,+}^*\right\}\right]=0,
\end{eqnarray}
where we in the first step used Eq.~(\ref{uv-relation}) (in the
negative energy bransch term) and in the second step used
Eq.~(\ref{BdGsym}). We therefore have the following current formula in
the semiconductor picture:
\begin{equation}
j=\frac{e\hbar}{m}\sum_{\beta}\sum_k\mbox{Im}
\left\{u_{k\beta}^*\nabla u_{k\beta}
+v_{k\beta}^*\nabla v_{k\beta}\right\}f(E_{k\beta}).
\label{currentop_semicond-pic}
\end{equation}

We may now turn the sum over hole-like quasiparticles into a sum over
electron-like quasiparticles by using the symmetry in Eq.
(\ref{BdGsym}) once again. The current formula will then be
\begin{eqnarray}
j=\frac{e\hbar}{m}\sum_{\beta}\sum_{k=k^e}\mbox{Im}
\left\{u_{k\beta}^*\nabla u_{k\beta}+
v_{k\beta}^*\nabla v_{k\beta}\right\}\nonumber\\
\times\left[2f_{FD}(E_{k\beta})-1\right],
\label{currentop_semi-e}
\end{eqnarray}
which is the one written down in Eq.~(\ref{currentdensity}). The
states included in Eq.~(\ref{currentop_semi-e}) is shown in
Fig.~\ref{fig11}(c).

In conclusion, we can use the completeness relation [Eq.
(\ref{uv-relation})] and the symmetry of the BdG equation [Eq.
(\ref{BdGsym})] to rewrite the formula for the current into different
forms. Which form to use is a matter of convenience as long as one
sums over the right set of scattering states. For instance in
Refs.~\onlinecite{Hurd}, \onlinecite{HDB2}, and \onlinecite{HDB1}
Eq.~(\ref{currentop_semi-pic}) was used; on the other hand, in this
paper and in Ref.~\onlinecite{LJHW} Eq.~(\ref{currentop_semi-e}) has
been used.

\section{Some remarks on the current}\label{App_C}

\subsection{Derivation of the symmetry $I(-V)=-I(V)$}
In this subsection we show the relation $I(-V)=-I(V)$ for both
perpendicular and parallel current density.

Letting $V\rightarrow-V$ and $E\rightarrow -E$ we see from
Eq.~(\ref{uv}) that the Andreev reflection amplitude $A_n\rightarrow
-A_n^*$.  Studying Eq.~(\ref{abc}), this gives
\begin{eqnarray}
\alpha_n&\rightarrow&\alpha_n^*,\nonumber\\
\beta_n&\rightarrow&\beta_n^*,\nonumber\\
\gamma_n&\rightarrow&\gamma_n^*,
\end{eqnarray}
implying that Eq.~(\ref{recd}) determining $d_n$ changes to
\begin{equation}
\alpha_n^*d_{n+2}+\beta_n^*d_n+\gamma_n^*d_{n-2}=rJ(E)\delta_{n,0}.
\end{equation}
After complex conjugation and multiplication of the factor $r/r^*$, we
have
\begin{equation}
(\alpha_nd_{n+2}^*+\beta_nd_n^*+\gamma_nd_{n-2}^*)\frac{r}{r^*}
=rJ(E)\delta_{n,0}.
\end{equation}
If we compare this recurrence equation to the one in Eq.~(\ref{recd}),
we see that
\begin{equation}
d_n\rightarrow d_n^* r/r^*.
\label{d-negV}
\end{equation}
Using Eq.~(\ref{d-negV}) in Eqs.~(\ref{ac_L}) and (\ref{b}), we get
\begin{eqnarray}
a_n&\rightarrow& a_n^*,\\
b_n&\rightarrow& -b_n^*,\\
c_n&\rightarrow& -c_n^* r/r^*,
\label{abc-negV}
\end{eqnarray}
which implies [from Eq.~(\ref{Teh})] that
\begin{equation}
T_x^{\tau}(E,\theta,m)\rightarrow [T_x^{\tau}(E,\theta,m)]^*.
\label{Tx-negV}
\end{equation}

A minus sign is extracted when we let $E\rightarrow-E$ in
Eq.~(\ref{AmBm}) since
\begin{equation}
\tanh(-E/2k_BT)\rightarrow-\tanh(-E/2k_BT).
\label{tanh-negV}
\end{equation}

Taking Eq.~(\ref{Tx-negV}) and Eq.~(\ref{tanh-negV}) into account we
get from Eq.~(\ref{AmBm})
\begin{equation}
A_{x,0}\rightarrow -A_{x,0}.
\end{equation}
Therefore, one finally finds
\begin{equation}
I_{x,0}(-V)=-I_{x,0}(V).
\end{equation}

The same kind of reasoning as above for the perpendicular case is used
to put the transformations Eqs.~(\ref{d-negV})-(\ref{abc-negV}) and
(\ref{tanh-negV}) into the corresponding equations for $j_{y,0}$.
Therefore, we also have
\begin{equation}
j_{y,0}(-V)=-j_{y,0}(V).
\end{equation}

\subsection{Disappearance of ac components}
Here we set out to show analytically how the ac components disappear
for some orientations of the superconducting order
parameters.\cite{LJHW} The argument presented here is not
$Z$-dependent: it is true for any barrier strength.

The keypoint is to compare current contributions associated with
injection angles $\pm\theta$ and see how they add up. We can separate
the integrand which we integrate over the angles
$\theta\epsilon[-\pi/2,\pi/2]$ into two functions: first a
trigonometric function and second a complicated function depending on
the gaps. The trigonometric function is $\cos\theta$ (even in
$\theta$) for the $x$ current and $\sin\theta$ (odd in $\theta$) for
the $y$ current. We will show that the second function is odd or even
in $\theta$ depending on the orientation of the superconductors.  The
combination of the two functions will therefore be odd in some cases
and therefore integrate to zero.

{\it Case 1: $d_0|d_0$ junction.} In this case the gaps are symmetric
in $\theta$ on both the left and the right side, implying that the
second function is even in $\theta$. Taking the trigonometric
functions into account, the $x$ current therefore contains all
components $m$ while the $y$ current is zero for all $m$.

{\it Case 2: $d_0|d_{\pi/4}$ junction.} This case is shown in
Fig.~\ref{fig2}. The gap on the right side is oriented so that the
negative lobe appears for negative angles:
$\Delta_2(-\theta)=-\Delta_2(\theta)=\Delta_2(\theta)\exp(i\pi)$.
This extra phase $\pi$ for the negative angles can be given to
$\phi_0$. As noted in Appendix~\ref{App_A}, the phase difference over
the junction $\phi_0$ is factored out in the beginning of the
calculation, but it reappears in the phase $\Omega_m$ in the final
expression for the current, see Eq.~(\ref{Icos_Ie}). We therefore have
\begin{equation}
\Omega_m(-\theta)=\Omega_m(\theta)+m\pi.
\label{symOmega}
\end{equation}
In addition, we have (taking also the trigonometric function into
account)
\begin{equation}
\left\{
\begin{array}{l}
A_{x,m}(-\theta)=A_{x,m}(\theta)\\
B_{x,m}(-\theta)=B_{x,m}(\theta)\\
A_{y,m}(-\theta)=-A_{y,m}(\theta)\\
B_{y,m}(-\theta)=-B_{y,m}(\theta)
\end{array}
\right.
\Rightarrow
\left\{
\begin{array}{l}
C_{x,m}(-\theta)=C_{x,m}(\theta)\\
\alpha_{x,m}(-\theta)=\alpha_{x,m}(\theta)\\
C_{y,m}(-\theta)=-C_{y,m}(\theta)\\
\alpha_{y,m}(-\theta)=\alpha_{y,m}(\theta)
\end{array}
\right.
\label{symABCA}
\end{equation}
Using Eq.~(\ref{symOmega}) and Eq.~(\ref{symABCA}), we can rewrite the
$x$ current
\begin{equation}
\frac{I_x}{\sigma_0}=\sum_m C_{x,m}(\theta>0)(1+e^{im\pi})
e^{i(\Omega_m-\alpha_{x,m})},
\label{symIx}
\end{equation}
and the $y$ current
\begin{equation}
\frac{I_y}{\sigma_0}=\sum_m C_{y,m}(\theta>0)(1-e^{im\pi})
e^{i(\Omega_m-\alpha_{y,m})},
\label{symIy}
\end{equation}
where the functions $C_{x/y,m}(\theta>0)$ contains an integration over
positive angles only. It is clear from Eq.~(\ref{symIx})
[Eq.~(\ref{symIy})] that the odd (even) components of the $x$ ($y$)
current will be zero for this orientation. Note that the $m=0$
component of the $y$ current is also zero.

{\it Case 3: $d_{\pi/4}|d_{\pi/4}$ junction.} Here we have an extra
phase $\pi$ on both the left and the right side for negative angles.
This means that we have no effective extra phase that we can factor
out together with $\phi_0$. Instead we see in Eq.~(\ref{uv}) that the
Andreev reflection amplitudes are
$A_{1/2}(-\theta)=-A_{1/2}(\theta)=A_{1/2}(\theta)\exp(i\pi)$.
Studying the equations in Appendix~\ref{App_A}, only combinations of
two $A$ multiplying each other appear. Thus the extra phases $\pi$
cancel. Taking also the trigonometric function into account, the
integrand will therefore be even (odd) for all $m$ components of the
$x$ ($y$) current in the same way as for the $d_0|d_0$ junction.

For orientations in between $\alpha_{1/2}=0$ and $\alpha_{1/2}=\pi/4$,
the cancellation effects will not be perfect and all components $m$
are non-zero for both the $x$ and $y$ currents.

\section{Analytic expressions for large and small voltages}\label{App_D}
\subsection{Large voltage in the ballistic limit}
In the large voltage limit ($eV>>\Delta_0$) we only need to consider
one Andreev reflection. This means that only coefficients $a_n$,
$b_n$, $c_n$, and $d_n$ with indices $n=0$ and $n=\pm2$ have to be
taken into account.

In the ballistic limit ($Z=0$), the surviving coefficients from an
electron-like quasiparticle injected from the left superconductor at
energy $E$ and angle $\theta$ will be $a_{0}^{\rightarrow}$ (the
injected electron) and $b_{2}^{\rightarrow}$ (the hole created from
Andreev reflection at the right superconductor).  As outlined in
Ref.~\onlinecite{HDB1} we can then perform the integration over energy
analytically in the limit $eV\rightarrow\infty$ and get the following
expressions for the currents
\begin{eqnarray}
\frac{I_{x,0}(V)}{\sigma_0}&=&\frac{eV}{\Delta_0}+
\frac{2}{3}\int_{-\pi/2}^{\pi/2}d\theta\cos\theta
\left[\frac{|\Delta_1(\theta)|}{\Delta_0}+
\frac{|\Delta_2(\theta)|}{\Delta_0}\right]\nonumber\\
&=&\frac{eV}{\Delta_0}+I_{excess},\nonumber\\
\frac{j_{y,0}}{(\sigma_0/L_y)}&=&\frac{2}{3}\int_{-\pi/2}^{\pi/2}
d\theta\sin\theta\left[\frac{|\Delta_1(\theta)|}{\Delta_0}+
\frac{|\Delta_2(\theta)|}{\Delta_0}\right],
\label{largeV_currents}
\end{eqnarray}
valid for large voltages. Note that both the excess current and the
surface current is independent of voltage when $eV>>\Delta_0$.

\subsection{Small voltage in the ballistic limit}
In the ballistic case there is a nonzero current contribution in the
small-voltage limit for both the perpendicular ($x$) and parallel
($y$) direction. For the ballistic one-channel SNS junction with
different $s$-wave gaps one has previously derived that the
perpendicular current is proportional to the smallest of the
gaps.\cite{HDB1} Generalizing this result to our case we have for the
current in the $x$-direction
\begin{equation}
\frac{I_x}{\sigma_0}=\int_{-\pi/2}^{\pi/2}d\theta\cos\theta
\mbox{min}(\frac{|\Delta_1(\theta)|}{\Delta_0},
\frac{|\Delta_2(\theta)|}{\Delta_0}).
\label{curr_x_smallV}
\end{equation}

In the same way we find for the parallel current density in the normal
region that
\begin{equation}
\frac{j_y}{(\sigma_0/L_y)}=\int_{-\pi/2}^{\pi/2}d\theta\sin\theta
\mbox{min}(\frac{|\Delta_1(\theta)|}{\Delta_0},\frac{|\Delta_2(\theta)|}
{\Delta_0}).
\label{curr_y_smallV}
\end{equation}

\end{document}